\documentclass[11pt,a4paper]{article}

\usepackage[british]{babel}

\usepackage[a4paper,top=2cm,bottom=2cm,left=2.5cm,right=2.5cm,marginparwidth=1.75cm]{geometry}

\usepackage[style=alphabetic, backend=biber]{biblatex} 
\DeclareLanguageMapping{british}{british-apa} 

\usepackage{amsmath}
\usepackage{multirow}
\usepackage{comment}
\usepackage{graphicx}
\usepackage[colorlinks=true, allcolors=black]{hyperref}
\usepackage{hyperref}
\usepackage{orcidlink}
\usepackage[title]{appendix}
\usepackage{mathrsfs}
\usepackage{amsfonts}
\usepackage{booktabs} 
\usepackage{caption}  
\usepackage{threeparttable} 
\usepackage{algorithm}
\usepackage{algorithmicx}
\usepackage{algpseudocode}
\usepackage{listings}
\usepackage{enumitem}
\usepackage{chngcntr}
\usepackage{booktabs}
\usepackage{lipsum}
\usepackage{subcaption}
\usepackage{authblk}
\usepackage[T1]{fontenc}    
\usepackage{csquotes}       
\usepackage{diagbox}
\usepackage{amsthm} 

\usepackage{setspace}
\usepackage{titlesec}
\titleformat{\section} 
  {\normalfont\Large\bfseries}{\thesection.}{1em}{}
  
\newcommand{\subsubsubsection}[1]{%
  \vspace{\baselineskip}
  \noindent\textbf{#1\\}\quad
}
\usepackage{float}   
\usepackage{caption} 


\newcommand\blfootnote[1]{%
  \begingroup
  \renewcommand\thefootnote{}\footnote{#1}%
  \addtocounter{footnote}{-1}%
  \endgroup
}

\title{On the robustness of Mann-Kendall tests used to forecast critical transitions}

\author[1,2,*]{Tristan Gamot}
\author[1,3,*]{Nils Thibeau{-}{-}Sutre}
\author[1]{Tom J.M. Van Dooren}
\affil[1]{\small Institut d’Ecologie et des Sciences de l’Environnement de Paris (iEES-Paris), Sorbonne Université, CNRS, INRAe, IRD, Université Paris Créteil, Université Paris cité, 75005 Paris, France}
\affil[2]{Centre de Physique Théorique (CPHT), CNRS, École polytechnique, Institut Polytechnique de Paris, 91120 Palaiseau, France}
\affil[3]{Aix Marseille Univ, CNRS, I2M, Marseille, France
}
\affil[*]{These two authors contributed equally to this work.}

\date{}  

\begin{document}
\maketitle

\begin{abstract}
Non-parametric approaches to test for trends in time series make use of the Mann-Kendall statistic. Based on asymptotic arguments, these tests assume that its distribution follows a Gaussian distribution, even for autocorrelated time series. Recent results on the lack of validity of this assumption urge a robustness analysis of these approaches. 
While the issue is relevant across a wide range of applications, we illustrate it here in the context of detecting \textit{early warning signals} (EWS) of critical transitions, which are used across a variety of research domains, and where commonly applied methods generate autocorrelation. 
We present a broad analysis, covering all types of critical transitions commonly investigated in EWS studies. We compare empirical distributions of the Mann-Kendall statistic computed from classical EWS indicators preceding critical transitions to the theoretical distributions hypothesized by Mann-Kendall tests. We detect mismatches leading to inflated type I error rates, which would routinely lead to announcing a critical transition while it is not occurring. In contrast to a recent recommendation, we conclude that the use of Mann-Kendall tests for trend detection in the context of forecasting critical transitions should be avoided. We point out several alternative methods available instead.
\end{abstract}

\vspace{2em}


\blfootnote{\footnotesize Corresponding author: Tristan Gamot, email: \texttt{tristan.gamot[at]ens-paris-saclay.fr}}

\newpage


\section{Introduction}

 Many complex systems exhibit critical transitions, also called tipping points, at which the system shifts abruptly from one state to another \cite{scheffer2009critical}. The term "tipping" is by now used in a very broad sense to denote qualitative changes in dynamical systems. So-called  "bifurcation-induced tipping" or "B-tipping" \cite{ashwin2012tipping} corresponds to qualitative changes in the underlying model of the dynamical system which are due to the crossing of a bifurcation point caused by a parameter change. Standard methods from fast–slow systems have been used to formalize this definition \cite{kuehn2011mathematical}. Systems approaching a bifurcation point show a characteristic behavior called \textit{critical slowing down}: following perturbations from a stable equilibrium, the relaxation time to the equilibrium increases as the bifurcation parameter approaches the bifurcation point \cite{wissel1984universal, kuehn2011mathematical, strogatz2018nonlinear}. In this way, the system's resilience to perturbations decreases, leading - among other effects - to increases in the variability of states visited and in the autocorrelation between successive states when approaching the bifurcation point.
In this context, \textit{early warning signals} (EWS) are statistical indicators of \textit{critical slowing down} which can be used to predict a critical transition before it is reached \cite{van2007slow, dakos2008slowing}. 
When the goal thus becomes detecting an increase in the variability or auto-correlation of the system state variables, the power of these tests should become maximized, while controlling the probability of false positives to a desired level.
Estimating these indicators in overlapping rolling windows superimposed on the original time series and then test for trends in the resulting indicator time series is the standard method \cite{dakos2012methods}. 

The detection of trends in time series is a common aim in many research fields and a number of parametric and non-parametric tests have been developed to this end. Among them, the non-parametric Mann-Kendall test \cite{mann1945nonparametric, kendall1975rank} is a classical test. As Kendall originally proved that Kendall's tau (normalized by its standard deviation) of i.i.d. (independent and identically distributed) random variables converges in law to a Gaussian distribution \cite{kendall1975rank}, the corresponding test assumes that the Mann-Kendall statistic is normally distributed. For time series of finite length, this approximation is already satisfactory for lengths above ten \cite{kendall1975rank}.  
However, there is no a priori reason why empirical time series would be samples of i.i.d. random variables. Moreover, in the case of time series of stochastic variables used as putative EWS, successive values cannot be independent. Indeed, the rolling windows used to calculate each value of the statistic overlap: most original observations will be used in several windows to calculate different values of the EWS indicator. 

By defining a corrected variance of the Mann-Kendall statistic, two main modifications of the Mann-Kendall test have been developed for the case of auto-correlated data \cite{hamed1998modified, yue2004mann}. The use of these modified Mann–Kendall tests to evaluate the significance of time trends in EWS indicators of critical transitions was recently recommended \cite{chen2022practical}, with guidelines. In fact, this test has been applied repeatedly in this context \cite{tong2014early, george2020early, wang2020early, bos2022anticipating, ismail2022early, helmich2023detecting, liu2024early, robinson2025assessing, jarvis2025early}.
However, the modified Mann-Kendall tests proceed under the assumption that distribution of the Mann-Kendall statistic is Gaussian. Yet, recent results \cite{gamot2026gaussian} have shown that its distribution is non-Gaussian in finite-length time series with sufficient autocorrelation.

This provides the rationale for this article: in the context of detecting trends for EWS of critical transitions, are Mann-Kendall tests adequate?
We briefly present the Mann-Kendall tests and introduce the methodology involved to calculate EWS of critical transitions. 
To account for the behaviour of these tests across a range of realistic data-generating processes, we consider the normal forms of codimension-one local bifurcations in continuous-time systems with several types of noise, thereby covering all types of critical transitions ("B-tipping") commonly investigated in EWS studies \cite{kuehn2011mathematical, o2018stochasticity}. 
We study the sensitivity of the tests to several variables, including the bifurcation type, the EWS indicator and the size of the rolling window used for estimation.
Based on the results from this study of robustness, we outline the limitations of the Mann-Kendall trend detection tests and, in particular, recommend the use of alternative tests in the case of EWS trend testing.

\section{Trend testing for \textit{early warning signals} of critical transitions} \label{sec:ews_trend_testing}

\subsection{Kendall's and Mann-Kendall's tau}
A time series $\left\{x_i\right\}$ of length $n$ can be viewed as a sample of $n$ random variables $\left\{X_i\right\}$, indexed by the time order of observations $i \in \left\{ 1,2,...,n \right\}$.  Here, we suppose that two different observations cannot have the same value, i.e. $x_i \neq x_j$ if $i \neq j$, which is realistic for real-world datasets. For cases with ties, consult \cite{kendall1975rank}.

Kendall's rank correlation coefficient, often called Kendall’s $\tau$, is a measure of rank correlation (also called concordance) between two time series $\left\{x_i\right\}_{i=1,...,n}$ and $\left\{y_i\right\}_{i=1,...,n}$. It was first introduced by Kendall \cite{kendall1938new} and is defined as: 
\begin{equation*}
    \tau =   \frac{1}{\binom{n}{2}} \sum_{1\leq i<j \leq n} \textrm{sgn}(x_j-x_i) \textrm{sgn}(y_j-y_i) 
\end{equation*}
where sgn(.) is the sign function. Note that Kendall's $\tau$ takes discrete values in the interval $[-1,+1]$.

If the $\left\{y_i\right\}_{i=1,...,n}$ values are replaced by the time order of the time series i.e. $i=1,2,\dots,n$, Kendall’s $\tau$ quantifies monotonic trends in the $\left\{x_i\right\}$ time series: 
\begin{equation*}
    \tau_{\text{\tiny MK}} =  \frac{1}{\binom{n}{2}} \sum_{1\leq i<j \leq n} \textrm{sgn}(x_j-x_i)
\end{equation*}

As this quantity is naturally associated with the Mann-Kendall trend tests, we propose to call it Mann-Kendall's tau and to denote it $\tau_{\text{\tiny MK}}$.

A $\tau_{\text{\tiny MK}}$ value of $+1$ (resp. $-1$) corresponds to a strictly increasing (resp. decreasing) time series. To ease notations, we refer to both the random variable as well as a realization as $\tau_{\text{\tiny MK}}$. This does not hinder understanding.

The original Mann-Kendall test assesses values of $\tau_{\text{\tiny MK}}$, assuming independence between all points in a time series \cite{mann1945nonparametric}. The rationale is that in this case, $\tau_{\text{\tiny MK}}$ normalized by its standard deviation - which is qualitatively the test statistic, see appendix \ref{annexe:kendall_tests} - converges in distribution to a Gaussian as the number of observations in the time series increases.

This test was extended by Hamed and Rao \cite{hamed1998modified} and Yue and Wang \cite{yue2004mann}, to account for autocorrelated data, and is now widely adopted in practice. As positive and negative autocorrelations increase vs. decrease the variance of Mann-Kendall's tau, these extensions proposed to a different normalization of the test statistic. Guided by asymptotic arguments, the distributions were still assumed to be Gaussian. See Appendix \ref{annexe:kendall_tests} for a detailed description of the different tests and their assumptions.

In many cases, however, the empirical distribution of the normalized Mann-Kendall tau is far from Gaussian when data are autocorrelated and time series are finite (see for example \cite{hamed2009exact}). Recently, a study found scaling laws which determine the conditions under which the Gaussian approximation remains valid for finite-length time series generated by stationary AR(1) and MA processes \cite{gamot2026gaussian}. The high robustness of $\tau_{\text{\tiny MK}}$ with regard to the data generating process (as it is based on data ranks, and not on data values) suggests that similar results might be found across a larger set of data generating processes, which would make the scaling laws more broadly relevant in the analysis of statistical indicators of critical transitions - the EWS. 

\subsection{\textit{Early warning signals} of bifurcation-induced tipping points}
\subsubsection{Data generating processes}\label{sec:bif_process}

Bifurcations are categorized according to the different types of changes that the structure of a phase portrait can undergo. Each class of local bifurcation, determined by local stability changes of invariant sets, can be described by a normal form \cite{strogatz2018nonlinear}. This means that all bifurcations of the same class are locally topologically equivalent to their normal form \cite{strogatz2018nonlinear}, and the dynamics near each stable branch resembles the dynamics of the normal form.
Therefore, rather than focusing on a specific - complex - model, we analyze data generated from the normal forms associated with each class of local bifurcation. This approach allows us to study generic models for several types of critical transitions commonly examined in EWS studies, a methodology already used in the EWS literature \cite{kuehn2011mathematical, boettiger2012quantifying, o2018stochasticity, bury2020detecting, bury2021deep, dylewsky2024early, donovan2024characterising}.

We focus on EWS which precede the most relevant codimension-one local bifurcations in continuous-time models (i.e., differential equations), except for the Hopf bifurcation, which is analogous to the pitchfork type in polar coordinates \cite{kuehn2011mathematical}. Table \ref{tab:bif_liste} lists the local bifurcations considered. A similar analysis can be carried out for discrete-time dynamical systems (i.e., recurrence relations). 

\begin{table}[]
\centering
\centerline{
\begin{tabular}{|c|c|c|}
\hline
Local bifurcation & Normal form & Linearized dynamics with noise \\ \hline Fold &    $\dot x =- r - x^2$         &  \\ \cline{1-2} Transcritical     &  $\dot x = r x - x^2$  & Ornstein-Uhlenbeck process \\ \cline{1-2} Pitchfork   &  \parbox{2.5cm}{\begin{align*}
 & \dot x =r x + \mu x^3 \\
  \text{with } \mu &>  0  \text{ (subcritical case)} \\
  \text{or } \mu < & 0  \text{ (supercritical case)}
  \end{align*}} &
  \\ \hline
\end{tabular}}
\caption{List of considered codimension-one local bifurcations in continuous time, as well as their normal forms and associated linearized dynamics in the vicinity of a stable equilibrium with an additive noise term. $x$ denotes the state variable (or vector) and $r$ the one-dimensional bifurcation parameter.} \label{tab:bif_liste}
\end{table}

Fold (or saddle-node) bifurcations are widely studied to model sharp transitions to contrasting states. This may concern abrupt transitions in climate science \cite{dakos2008slowing} or in ecological models \cite{may1977thresholds}. Many hysteretic processes are modeled with systems involving two fold bifurcations. Near each stable branch, EWS have been applied to time trajectories of such scenarios (for example, \cite{dakos2012methods}). Transcritical bifurcations are frequently considered to model cases where a population goes to extinction when a parameter crosses a threshold, and EWS are used to anticipate such cases in ecology \cite{drake2010early} or infectious disease epidemiology \cite{southall2021early}.  Finally, the pitchfork bifurcation is used in climate science models, for example to model the switching between two simultaneously stable climatic regimes or atmospheric states \cite{nicolis2014dynamical}.

The normal forms (\ref{tab:bif_liste}) are deterministic and therefore need an additional stochastic part to model random perturbations and collect sample statistics. It can be introduced in different ways. Common choices are additive or multiplicative noise, often given without justification, even when contrasting choices are made for similar underlying deterministic models (see, for example, \cite{dakos2012methods} vs. \cite{chen2022practical}). Here we consider the case of additive noise: 
\begin{equation}\label{eq:EDS}
    \mathrm{d}x=f(x,r)\mathrm{d}t + \sigma \mathrm{d}W ,
\end{equation} 
where $f$ is the normal form of the local bifurcations presented in table \ref{tab:bif_liste}, parameterized by the bifurcation parameter $r$. Parameter $\sigma$ is independent of $x$, and $W$ is a Wiener process (Brownian motion). That is, $W(t) \sim \mathcal N(0,t) $ and $dW(t) \sim \mathcal{N}(0,dt)$ \cite{oksendal2003stochastic}. 
Appendix \ref{an:multiplicative} presents results for multiplicative noise.

\subsubsection{Choice of null model}\label{sec:null_model}
The primary pair of hypotheses when forecasting critical transitions on the basis of EWS can be formulated as follows: the system is not undergoing a bifurcation (null H0) versus the system is undergoing a bifurcation (alternative H1) \cite{dakos2012methods}, where we aim to reject H0. A significant trend in one of the EWS indicators is typically interpreted as strong evidence that the system is approaching a bifurcation point, due to the phenomenon of \textit{critical slowing down}.
In practice, many studies adopt a restricted null model in which the bifurcation parameter remains fixed over time \cite{southall2021early}. In the following manuscript, we adopt this steady-state definition of the null model as it seems the most common approach in this field. Although easier to implement and derive distributions of test statistics from, this null model can generate different distributions of statistics than the original null model of no bifurcation.

\begin{figure}[h]
    \centering
    \begin{subfigure}[b]{0.49\textwidth}
         \centering
         \includegraphics[width=1.\textwidth]{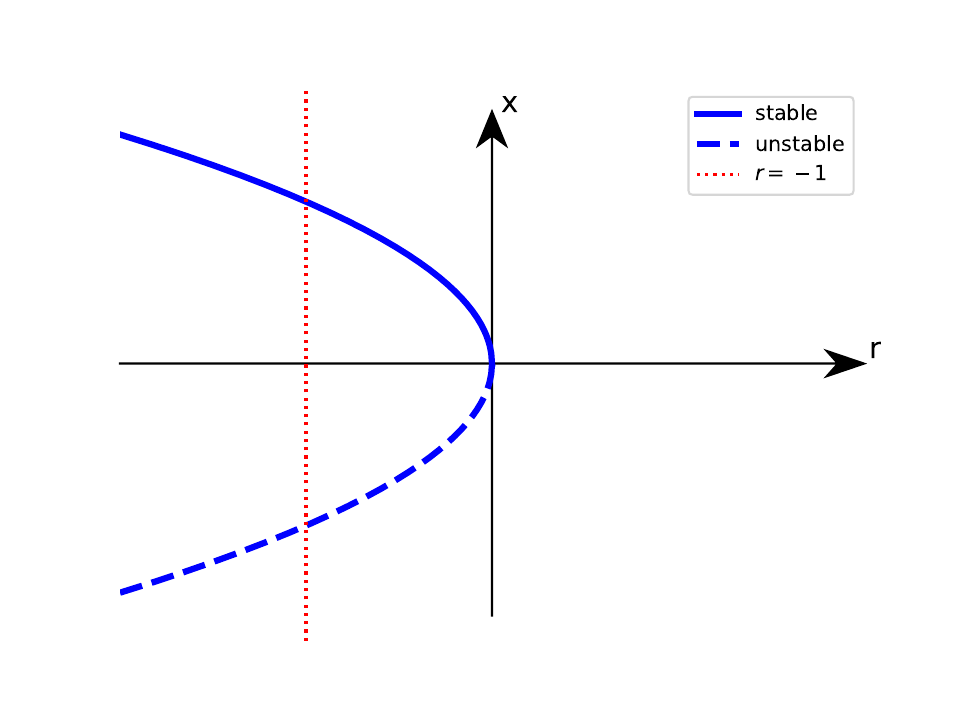}
        \caption{Fold (or saddle-node) bifurcation}
    \end{subfigure}%
    ~
    \begin{subfigure}[b]{0.49\textwidth}
        \includegraphics[width=1.\textwidth]{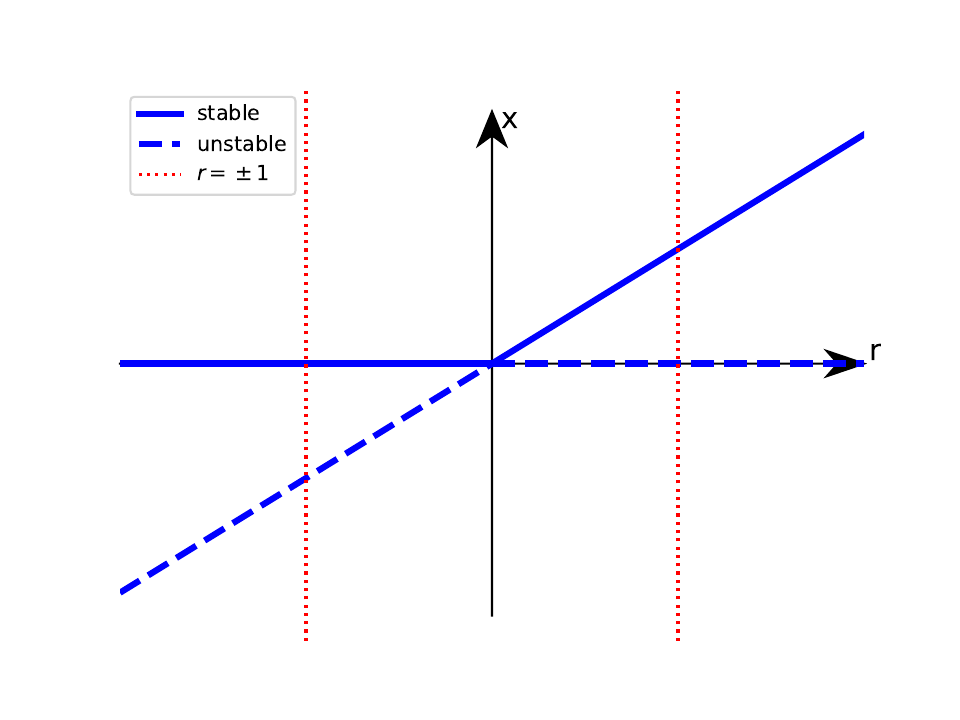}
        \caption{Transcritical bifurcation}
    \end{subfigure}%

    \begin{subfigure}[b]{0.49\textwidth}
         \centering
         \includegraphics[width=1.\textwidth]{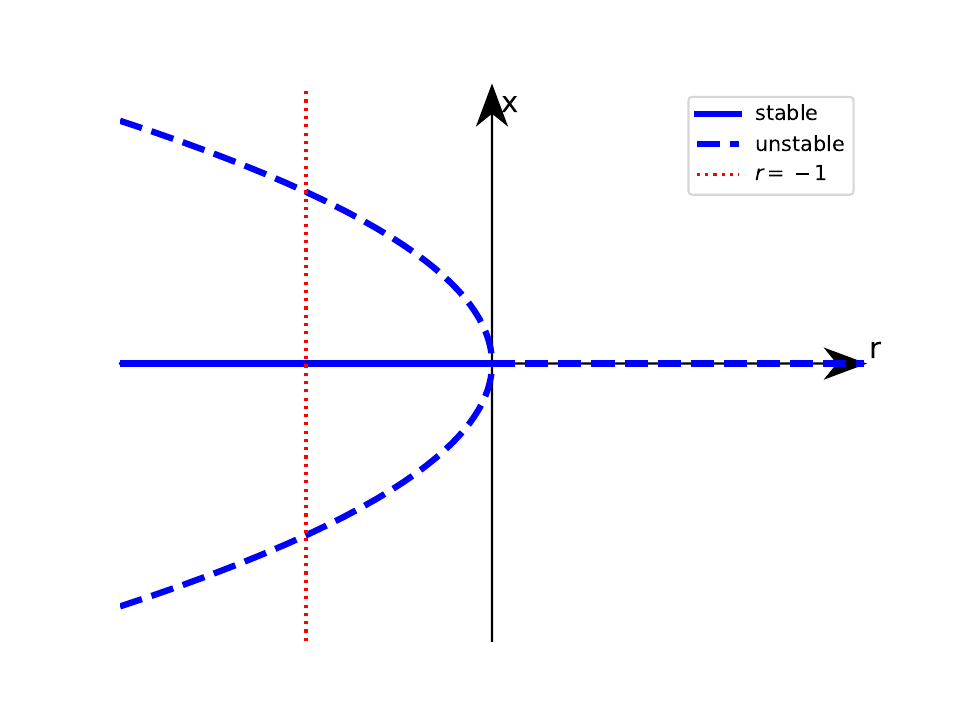}
        \caption{Subcritical pitchfork bifurcation}
    \end{subfigure}%
    ~
    \begin{subfigure}[b]{0.49\textwidth}
        \includegraphics[width=1.\textwidth]{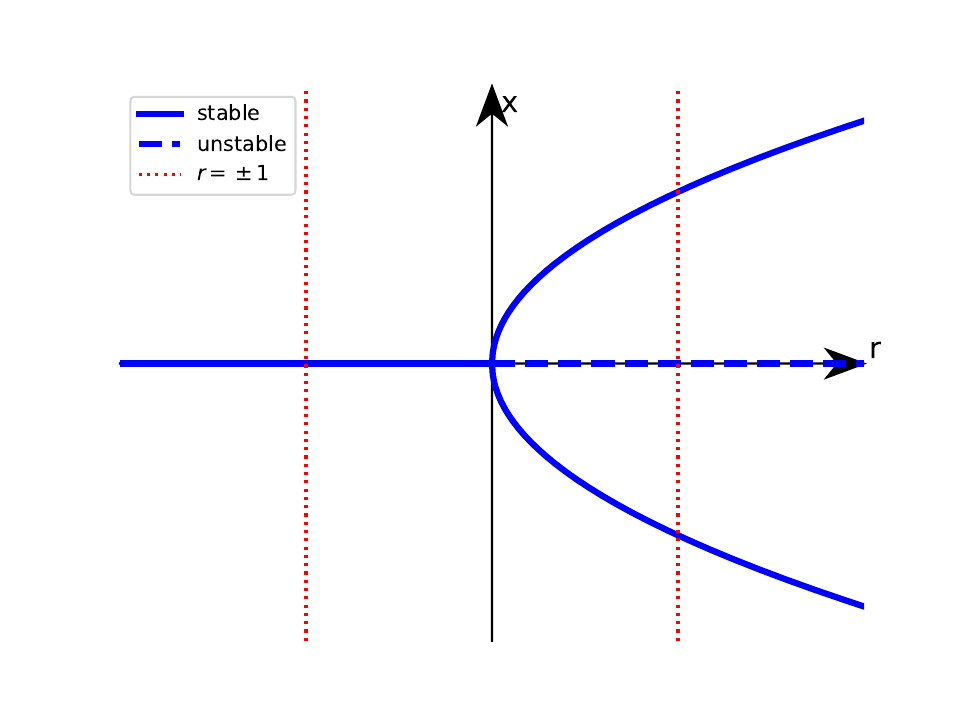}
        \caption{Supercritical pitchfork bifurcation}
    \end{subfigure}
    \caption{Bifurcation diagrams of the studied codimension-one local bifurcations. $x$ is the state variable and $r$ the bifurcation parameter, with the same notations as in Table \ref{tab:bif_liste}. Dotted red lines show the values of $r$ chosen for the steady-state null models.}
    \label{fig:bifurcations}
\end{figure}

We generate time series from the steady-state null model (fixed bifurcation parameter for the local bifurcations listed in Table \ref{tab:bif_liste}) in the vicinity of a stable equilibrium, with additive noise. These cases are illustrated in Figure \ref{fig:bifurcations}. Numerical integration of the resulting stochastic differential equations is performed  using the Python package \textit{sdeint} \cite{sdeint} and the stochastic Runge-Kutta algorithm SRI2 \cite{rossler2010runge}. After a transient period, the trajectory of the stochastic process is sampled uniformly to obtain what we refer to as the original time series in the remainder of this paper. This mimics time series of interest, which could arise from a variety of underlying models and are typically analyzed using EWS assuming the null hypothesis.

Equation \eqref{eq:EDS} can be linearized close to its stable equilibrium at a fixed value of the bifurcation parameter $r$, leading to an Ornstein-Uhlenbeck process. This approach is commonly used to derive analytical approximations for the behavior of EWS indicators depending on $\lvert r\rvert$, the distance between the actual value of the bifurcation parameter and the critical value at which the local bifurcation occurs \cite{kuehn2011mathematical, o2018stochasticity,bury2020detecting}. 
Since the Ornstein-Uhlenbeck process arises for all considered bifurcations - albeit with different linearization parameters - we expect the null distributions of the Mann-Kendall tau to be independent of the bifurcation type.
Note that Kuehn \cite{kuehn2011mathematical} has studied the range of validity of the linearization approximation, depending on the normal form. For curious readers, also note that the linearization parameters have different dependencies on the bifurcation parameter for each type of noise \cite{o2018stochasticity}. Yet, as our null model is for fixed $r$, this does not affect our approach.

\subsubsection{Estimation of \textit{Early Warning Signals}}
As the goal is to detect a decrease of resilience of the stable state as the bifurcation parameter approaches the bifurcation point, corresponding to \textit{critical slowing down}, one wants to test that EWS indicators exhibit a rising or falling trend prior to the critical transition. To monitor their time evolution, the standard method is to estimate these indicators in overlapping rolling windows superimposed on the original time series and then to test for trends in the time series of the estimated indicators \cite{dakos2012methods}. 
Non-stationarities in the mean can cause spurious indications of forthcoming transitions \cite{dakos2012methods}, for example due to parameter changes without bifurcation,  and therefore the original time series is usually detrended. We used Gaussian detrending with a 10\% bandwidth  \cite{dakos2012methods}.
There is no golden rule for choosing the right size of the rolling window given a finite time series, as there is a trade-off between the precision of the estimated indicators which increases with window length and the number of estimates available to detect a trend, which decreases with windows length. A commonly accepted, yet arbitrary, value for the window size is 50\% of the size of the dataset combined with a stride of one data point \cite{dakos2012methods}.
Several articles checked the effect of the window size in a sensitivity analysis, for values ranging from 5\% to 90\% \cite{lenton2011early, lenton2012early, kaur2020anticipating, southall2021early, chen2022practical}, concluding that it clearly influences the Mann-Kendall's tau values.

Thus, starting from an original time series $\{ x_i \}_{i=1,...,N}$ of length $N$, one can choose a rolling window size $q \in \{1,...,N\}$. Then, the EWS indicators are computed over rolling windows with a stride of one, leading to an EWS time series of length $n = N-q+1 = N(1-\alpha) +1$ where $\alpha = \frac{q}{N} \in [0,1]$ is the relative window size. The two most commonly used EWS indicators are the variance ${s^2}_i = \frac{1}{q-1} \sum^{i+q-1}_{k=i} (x_k-\Bar{x}_i)^2$, and the auto-correlation at lag-1 $\rho_1(i) = \frac{1}{q} \sum^{i+q-2}_{k=i} \frac{(x_{k+1} - \Bar{x}_i)(x_{k} - \Bar{x}_i)}{s^2_i}$ where $\Bar{x}_i$ is the mean of the time series over the i-th rolling window. See \cite{dakos2012methods} for a review of other indicators.

Hypotheses on trends are then tested on the EWS indicator time series. There are various ways of testing for trends, such as the original or modified versions of the Mann-Kendall test presented previously. The rolling window method introduces auto-correlation in the indicators because windows overlap (the same original time point will be used in several rolling windows). It seems more reasonable to apply the modified tests for autocorrelated data, which have recently been put forward by Chen et al. \cite{chen2022practical}.
Yet, as we have noted, these tests rely on assumed Gaussian distributions of the Mann-Kendall tau. In the next section, we will show that this assumption is invalid in general, thereby often rendering these tests anticonservative due to inflated type I error rates.

\section{Robustness of Mann-Kendall tests assuming Gaussian distributions}\label{sec:consequences}
\subsection{Relative insensitivity of the distribution of Mann-Kendall's tau to the data generation process and choice of indicator}

\begin{figure}
    \centering
    
    \begin{subfigure}[b]{.38\textwidth}
         \centering
         \includegraphics[width= 1.\textwidth]{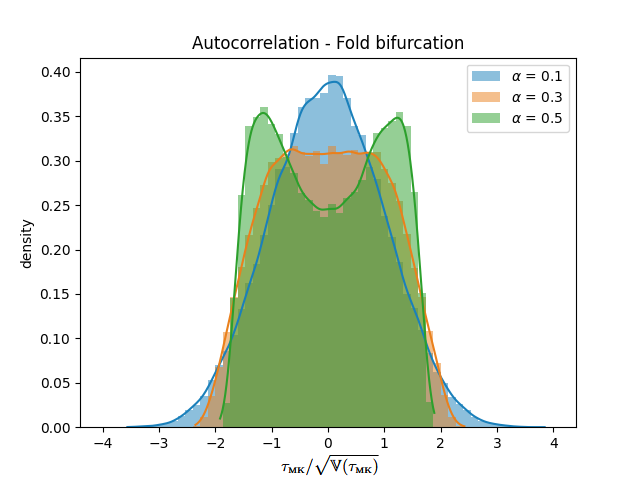}
     \end{subfigure}%
    ~
    \begin{subfigure}[b]{.38\textwidth}
         \centering
         \includegraphics[width= 1.\textwidth]{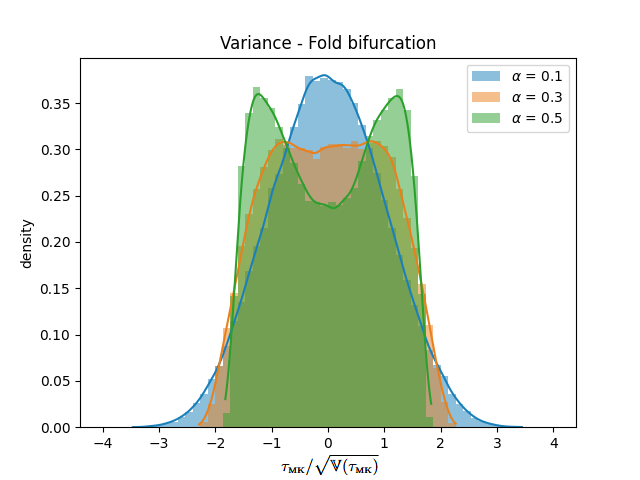}
     \end{subfigure}%

     \begin{subfigure}[b]{.38\textwidth}
         \centering
         \includegraphics[width= 1.\textwidth]{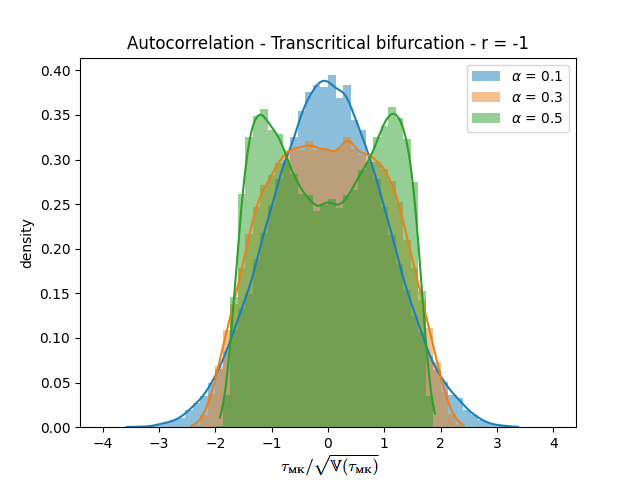}
     \end{subfigure}%
    ~
    \begin{subfigure}[b]{.38\textwidth}
         \centering
         \includegraphics[width= 1.\textwidth]{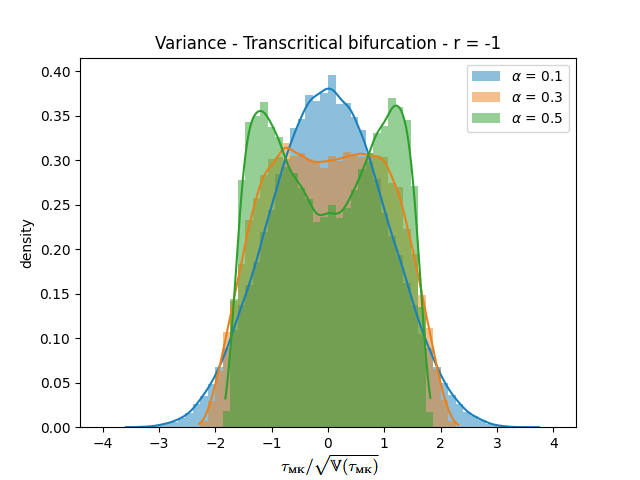}
         \end{subfigure}%

     \begin{subfigure}[b]{.38\textwidth}
         \centering
         \includegraphics[width= 1.\textwidth]{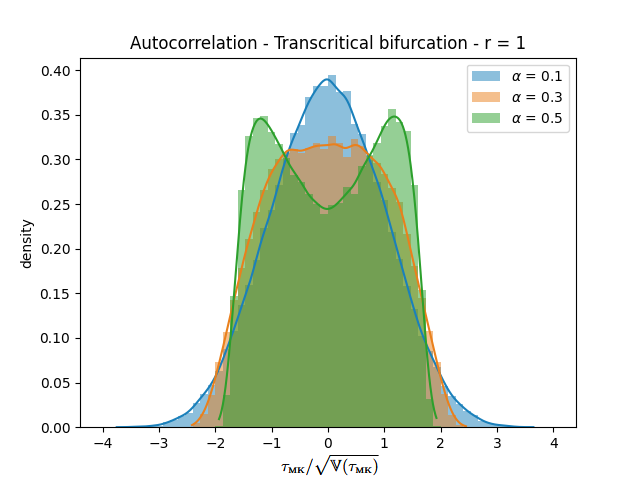}
     \end{subfigure}
    ~
    \begin{subfigure}[b]{.38\textwidth}
         \centering
         \includegraphics[width= 1.\textwidth]{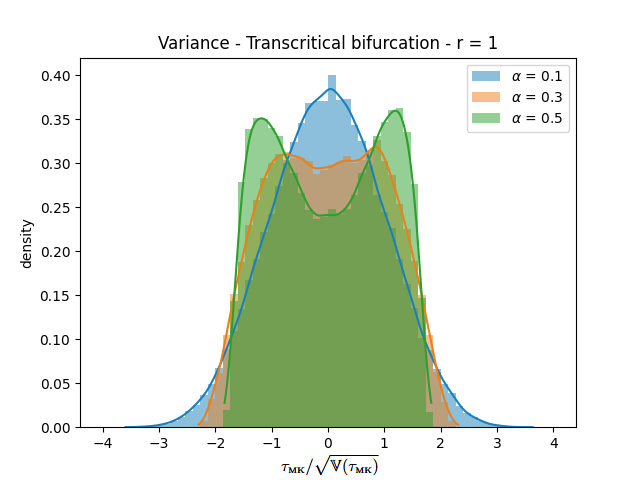}
     \end{subfigure}

     \begin{subfigure}[b]{.38\textwidth}
         \centering
         \includegraphics[width= 1.\textwidth]{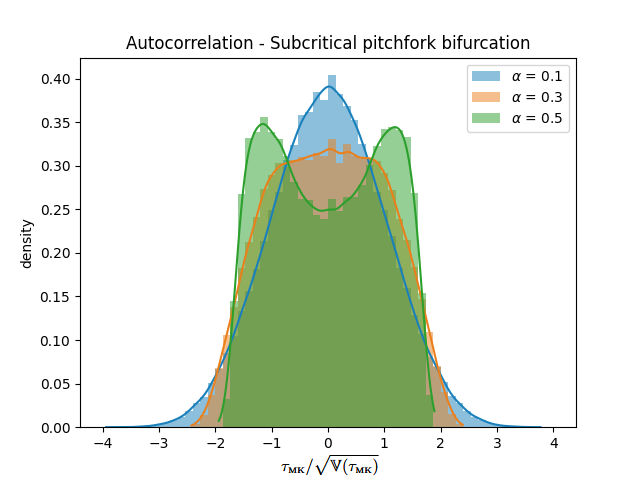}
     \end{subfigure}
    ~
    \begin{subfigure}[b]{.38\textwidth}
         \centering
         \includegraphics[width= 1.\textwidth]{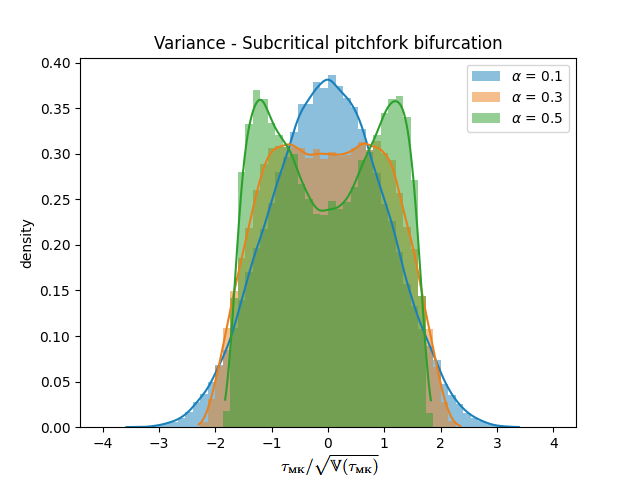}
     \end{subfigure}

    \begin{subfigure}[b]{.38\textwidth}
         \centering
         \includegraphics[width= 1.\textwidth]{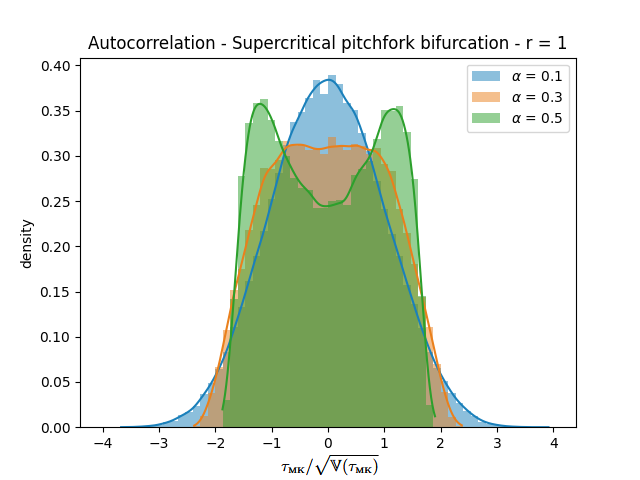} 
     \end{subfigure}%
    ~
    \begin{subfigure}[b]{.38\textwidth}
         \centering
         \includegraphics[width= 1.\textwidth]{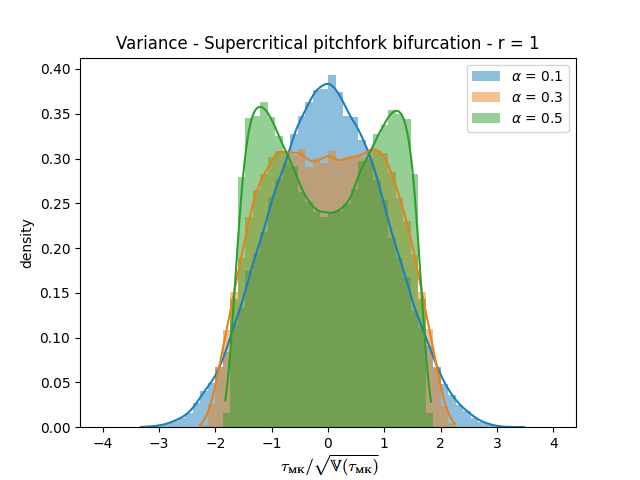} 
     \end{subfigure}

    \caption{Empirical distributions of normalized Mann-Kendall's tau calculated from lag-1 autocorrelation coefficients (left column) or variances (right column) over rolling windows of relative size $\alpha$ for original time series of length $N = 100$. Each row is generated for one of the normal forms listed in Table \ref{tab:bif_liste}.}
    \label{fig:similar_distributions}
\end{figure}

Figure \ref{fig:similar_distributions} presents empirical distributions of normalized Mann-Kendall's tau calculated from lag-1 autocorrelation coefficients and variances over rolling windows of  different relative sizes $\alpha$ for a fixed time series length. Time series were sampled from simulations of each normal form presented in Figure \ref{fig:bifurcations}, with the bifurcation parameter $r$ kept fixed for each time series at the values indicated in that figure per normal form (stationary null model). 

A first interesting result is that the empirical distributions of the statistics are strongly influenced by $\alpha$, with minimal sensitivity to the bifurcation type or the EWS indicator chosen. This suggests that the shape of the empirical distribution is mostly to the use of rolling windows and that parameter $\alpha$ can be used to predict its approximate distribution well.
The fact that the bifurcation type is not influencing the distribution of Mann-Kendall's tau is coherent with the fact that the bifurcations studied here exhibit similar linearized dynamics close to a stable equilibrium, as explained in Section \ref{sec:null_model}. 
Note that the length $N$ of the original time series also has a minimal influence on the distribution, see appendix \ref{an:length}. 

Furthermore, we observe that, as $\alpha$ increases, the probability density near zero decreases. For sufficiently large $\alpha$, the empirical distribution becomes bimodal and very different from a Gaussian.
This bimodality has been visible in results of several publications, in statistics obtained from simulated datasets produced by very different models. Table \ref{tab:bimodal_distrib} gives a non-exhaustive list of them.

\begin{table}[h]
\centering
\begin{tabular}{|c|c|c|}
\hline
Article & Figure & $\alpha $ \\ \hline
SI of \cite{dakos2008slowing}  &    S3    &   0.5    \\ \hline
\cite{dakos2012methods}     &    11    &   0.5    \\ \hline
\cite{chen2022practical}       &    5c    &   0.5    \\ \hline
\cite{boulton2015slowing}       &    1 \& 2    &   0.5    \\ \hline
\cite{thomas2015early}      &    5    &   0.5    \\ \hline
\cite{diks2019critical}        &     7   &    0.5   \\ \hline
\cite{shalalfeh2016kendall}       &    4    &   0.6    \\ \hline
\cite{deb2022identifying}     &   5     &   0.6    \\ \hline
\end{tabular}
\caption{List of figures in EWS studies where the bimodality of distributions of Mann-Kendall's tau is visible. The relative size $\alpha$ of the rolling windows used for EWS estimation in each study is given.}\label{tab:bimodal_distrib}
\end{table}

\subsection{Comparison of empirical null distributions with hypothesized distributions of Mann-Kendall tests}

\begin{figure}
     \centering
        \includegraphics[width= 1.\textwidth]{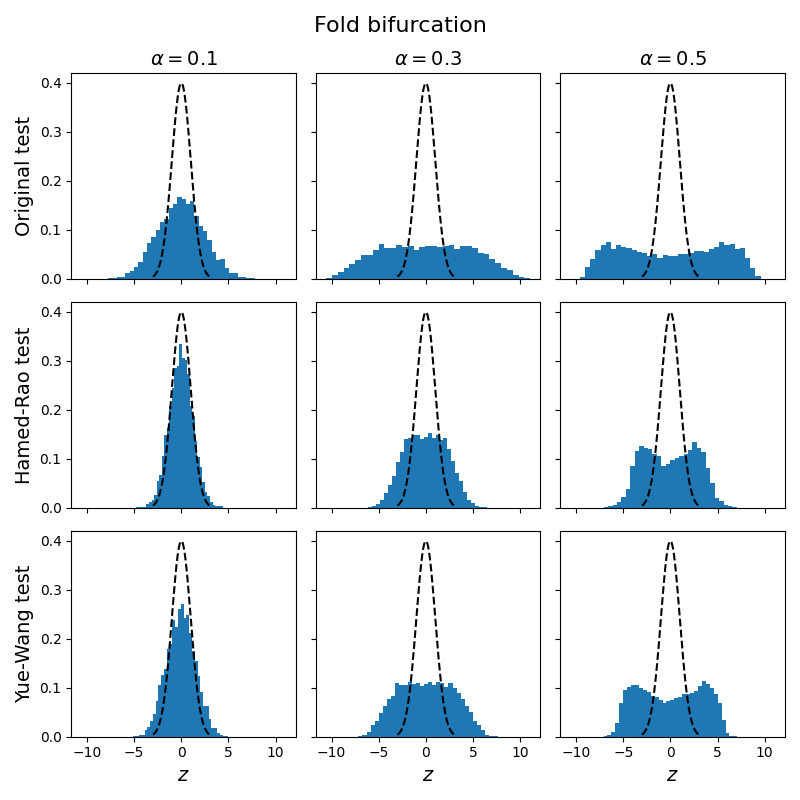}
        \caption{Comparison of simulated empirical distributions of test statistics $Z$ of the original Mann-Kendall test (first row) the Hamed and Rao test (second row) and the Yue and Wang test (third row) with a standard normal distribution, the distribution hypothesized by the tests. The $Z$ statistic is Mann-Kendall's tau renormalized differently for each test, see appendix \ref{annexe:kendall_tests}, and computed for the lag-1 autocorrelation coefficient time series estimated over rolling windows of relative size $\alpha$. The original time series, of length $N = 100$, are sampled from the fold bifurcation with fixed bifurcation parameter $r=-1$ (stationary null model).
        The two bottom rows show that the modified tests improve the match between empirical and theoretical null distributions, despite discrepancies. The case $N = 1000$ is shown in Figure \ref{fig:an_N_sample}.}
    \label{fig:distrib_X_rolling_windows}
     \end{figure}

As the distributions are primarily governed by $\alpha$, we can thus restrict ourselves in the following on time series generated by one bifurcation model, the fold, and one indicator, the autocorrelation coefficient at lag one. Figure \ref{fig:distrib_X_rolling_windows} compares empirical distributions of the normalized Mann-Kendall's tau of autocorrelation coefficients in simulations with the corresponding theoretical distributions used by the original test, as well as distributions of the two modified tests. 

Even though the modified tests are more adequate than the original one for small values of $\alpha$, we see that the theoretical distributions used by the tests are very poor approximations of the empirical null distributions in many cases. The renormalization of Hamed and Rao \cite{hamed1998modified} appears to be the least inaccurate among the three approaches.

As $\alpha$ increases, the discrepancy between empirical and theoretical null distributions increases. This arises from the fact that the rolling window method introduces auto-correlation between the estimated indicators on length scales of the order of the window size. This leads to inflated type I errors.

\subsection{Type I error rates}
\begin{figure}
     \centering
     \includegraphics[width=0.7\textwidth]{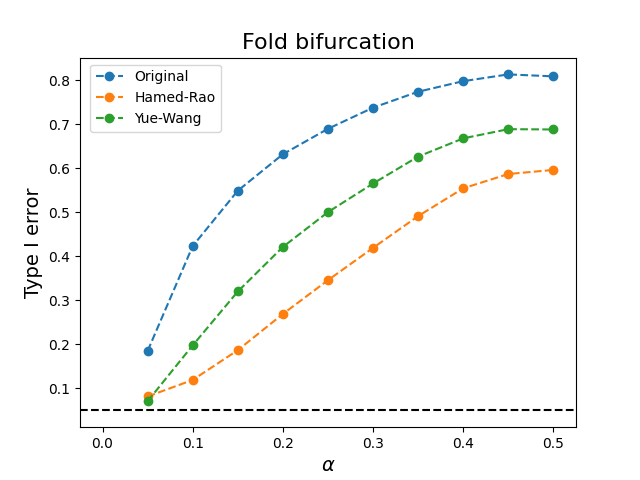}
    \caption{Empirical rejection rates of tests of the null hypothesis of no trend when the nominal 5\% significance level is applied. Recommended lag coefficients are taken into account to compute the auto-correlation functions required to estimate the variance of the theoretical Gaussian distributions (one for the Yue and Wang modified test and three for the Hamed and Rao modified test). The horizontal black line indicates the nominal 5\% significance threshold. Each original time series has length $N = 100$ and is simulated from the fold normal form (with fixed bifurcation parameter $r=-1$). The lag-1 autocorrelation (the chosen EWS indicator) is computed using rolling windows of relative size $\alpha$ with stride $1$. The figure uses results computed using the pymannkendall package \cite{hussain2019pymannkendall}. }
    \label{fig:window_test}
\end{figure}

In hypothesis testing, it is important to determine the proportion of false positive tests (type I error rate). Here this is the rate of rejecting the null hypothesis of no trend while it is true. 
Figure \ref{fig:window_test} presents empirical Type I error rates of the original and the two modified Mann-Kendall tests, for different relative window sizes and the fold bifurcation. Other EWS indicators and other bifurcation types investigated exhibit similar empirical distributions of Mann-Kendall's tau . 
For each test, the type I error rate is an increasing function of $\alpha$. It largely exceeds the nominal 5\% significance level, even for $\alpha = 0.05$. We expect the type I error rate to converge towards the 5\% level as $\alpha$ goes to zero.

This monotonicity is consistent with the observations that the difference between the hypothesized and true distributions increases as $\alpha$ increases, see Figure \ref{fig:distrib_X_rolling_windows}. This highlights that these tests are not very robust to the assumption of a Gaussian distribution under the null hypothesis. A sensitivity analysis to many parameters (type and intensity of noise, sample size, distance to the bifurcation) is available in the Appendix \ref{an:sensitivity_analysis}, confirming that this result holds true for a vast range of parameters values. 

In the case of EWS trend detection, where the recommendation was to take a relative window size of $50\%$ \cite{dakos2012methods}, the type I error is above $50\%$. 
This shows that the modified Mann-Kendall statistical tests are not suitable for trend detection in this case, and would routinely lead to "cry wolf" effects: announcing a critical transition while it is not occurring.  

\subsection{Ways to reduce the artificial autocorrelation induced by the rolling window method}

The distributions shown above seem very unsensitive to many parameters except the relative window size $\alpha$ of the rolling windows, suggesting that this method impedes the Mann-Kendall tests to be reliable.
However, one can estimate EWS on less overlapping rolling windows, so that successive EWS indicator values become less autocorrelated compared to more strongly overlapping windows. 

Therefore, we estimated EWS indicators using rolling windows with varying strides: 1 data point (i.e., fully overlapping windows, the conventional choice), strides at 2\%, and 5\% of the time series length, thereby progressively reducing the degree of overlap. For example, the overlap for a stride of 5\% and a relative window size of 10\% is 50\%, i.e., half of the data points used to estimate an EWS indicator in a window will be used in the following window.

\begin{figure}
     \centering
     \includegraphics[width=0.7\textwidth]{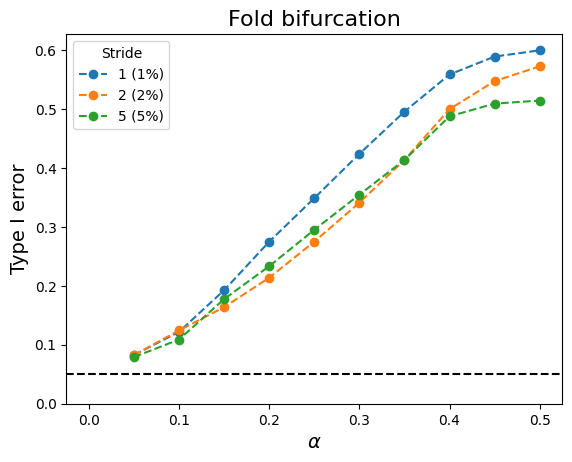}
    \caption{Empirical rejection rates of the null hypothesis of no trend for the Hamed and Rao modified Mann-Kendall test used with a 5\% significance level, as a function of the stride between successive rolling windows of relative size $\alpha$. The horizontal black line indicates the nominal 5\% significance threshold. Each original time series has length $N = 100$ and is simulated from the fold normal form (with fixed bifurcation parameter $r=-1$). The lag-1 autocorrelation (the chosen EWS indicator) is computed using rolling windows with corresponding stride.}
    \label{fig:overlap}
\end{figure}

Figure \ref{fig:overlap} shows the rejection rates of the null hypothesis of no trend for these configurations. We observe that as the overlap decreases (i.e., as the stride increases), type I error rates also decrease. However, the rejection rates remain substantially higher than the expected 5\% level. Moreover, using partially overlapping windows significantly shortens the length of the EWS indicator time series, which may reduce the power of trend analysis, especially in cases where only limited data are available. For example, using a stride of $5\%$ and a relative window size of $50\%$ leads to an EWS time series of length $n = 11$, for whatever length of the original time series. 

\section{Discussion}
We have observed that the null distributions of Mann-Kendall's tau of the two main EWS indicators depend mainly on the relative size of the rolling window used and not on the specific indicator nor bifurcation type.
For often used relative rolling window sizes, the distributions hypothesized by the Mann-Kendall tests do not give a good approximation of the empirical distribution. 
This is due to two reasons: firstly for large enough $\alpha$ the Gaussian shape is not appropriate, see Figure \ref{fig:similar_distributions}, and secondly the variances of the hypothesized Gaussian distributions of the tests are also not matching, see Figure \ref{fig:distrib_X_rolling_windows}. This is clearly visible for the case $\alpha = 0.1$, for which the empirical distribution is almost Gaussian in Figure \ref{fig:similar_distributions} but the hypothesized distributions of the tests have a larger variance in Figure \ref{fig:distrib_X_rolling_windows}. This second effect is amplified as the size of the original time series increases, see Figure \ref{fig:an_N_sample} from Appendix \ref{an:length}.
Note that the renormalizations proposed by the two modified tests do improve the match between empirical and theoretical null distributions of the test statistics.

The larger the size of the rolling window, the more likely it becomes to observe extreme Mann–Kendall's tau values compared to what is expected under the assumptions of the tests. This can misleadingly suggest the presence of a trend, thereby inflating the type I error rates of the tests. When we compared the rejection rates of the null hypothesis of no trend at the statistical significance level of 5\% for the Mann-Kendall tests, these depended on the value of the relative window size. These results show that the original and modified tests have rejection rates of the null hypothesis which are far too high, even for small relative window sizes.

Furthermore, even when the autocorrelation introduced by the rolling window procedure is reduced by decreasing the degree of window overlap, the type I error rates remain well above the expected 5\% nominal level. The sensitivity analysis presented in Appendix \ref{an:sensitivity_analysis} consistently shows that Mann-Kendall tests yield inflated type I error rates. Another approach to reduce the autocorrelation introduced by the rolling window procedure could consist in estimating EWS indicators using rolling windows in which data are weighted with Gaussian kernels, assigning more weight to points near the center of the window. This reduces the "effective" window's size. To the best of our knowledge, this method has not yet been applied in the EWS literature. It requires a separate investigation to assess the consequences of this approach.

Since autocorrelation in the original time series could bias the outcome of modified Mann-Kendall tests, we investigated the effects of that in Appendix \ref{an:autocorr}. In our normal form framework of critical transitions, the $r$ bifurcation parameter controls the autocorrelation in the original time series. These become less and less autocorrelated as $r$ increases, but the type I error rates still substantially exceed the nominal 5\% level, see Figure \ref{fig:Appendix_Distance}, because of the overlapping rolling windows methodology.

Our robustness analysis challenges the use of non-parametric tests for trend detection in the context of EWS of critical transitions. For instance, Chen et al. \cite{chen2022practical} compared empirical distributions of the Mann-Kendall tau - computed from rolling-window estimates of EWS indicators (specifically, the variance) - to the theoretical distributions of both the classical Mann-Kendall test and the modified version proposed by Hamed and Rao \cite{hamed1998modified}.
Because the variance of the Gaussian approximation in the modified test is scaled to reflect the autocorrelation in the data, the authors concluded that the modified test adequately captures the empirical distribution. Based on this, they recommend using the modified Mann-Kendall test to assess the significance of observed tau values, highlighting its computational efficiency relative to bootstrap-based approaches.
While it is true that the variance of the modified Gaussian distribution matches the empirical variance more closely than that of the classical test, their figures (see 3, 5, and 6) clearly show that the Gaussian approximation itself is not a good fit. The empirical distributions they observe closely resemble those we present in Figure \ref{fig:distrib_X_rolling_windows}, and appear to depend primarily on the relative length of the rolling window. This is consistent with our findings.
Moreover, the robustness of distributions with respect to the underlying generative process is also evident when Chen et al. \cite{chen2022practical} is compared with this study: they analyze time series from a harvesting model, which differs significantly from our simulations based on normal forms of bifurcations with additive noise, yet the resulting empirical distributions exhibit the same patterns.

Thus, we arrive at the opposite conclusion to that of Chen et al. \cite{chen2022practical}, namely that existing Mann-Kendall tests are not suitable to assess the significance of Mann-Kendall's tau when testing for critical transitions. To our knowledge, we see two main alternative approaches to assess the significance that the system is undergoing a critical transition: use Mann-Kendall's tau on EWS indicator time series but with alternative tests that make no assumption on its distribution, or use methods to forecast critical transitions that do not use Mann-Kendall's tau to quantify trends, of which there are several. 

In the first case, an alternative approach is to use parametric tests based on resampling to study the significance of the Mann-Kendall tau of EWS time series. 
These tests rely on a specified null model that generates surrogate datasets with the same correlation structure and probability distribution as the original dataset, but under the assumption of stationarity. Common choices include autoregressive processes such as AR(1), or more generally, the best-fitting autoregressive moving average (ARMA(p,q)) given the original dataset \cite{dakos2008slowing}.
Surrogate time series are then simulated from this stationary process to estimate the probability of observing a Mann-Kendall tau as extreme as the one computed from the original time series, under the null hypothesis that the system follows the chosen stationary dynamics. These parametric tests are among the most widely used methods for assessing the statistical significance of observed trends in the EWS literature, likely because they rely on the most explicit hypothesis \cite{boettiger2012quantifying}.

The second class of methods does not rely on the trend of EWS indicators to assess the probability of a critical transition and therefore does not use Mann-Kendall's tau. Several metric-based indicators fall into this category such as the Brock–Dechert–Scheinkman test and conditional heteroskedasticity. Both have been reviewed in the context of EWS for critical transitions \cite{dakos2012methods}, although they are rarely applied in practice.
Other approaches involve model-based EWS indicators, in which data are fitted to specific models. As a result, the conclusions depend on the chosen model structure. For instance, one may compare two models that can each be transformed into the normal form of the underlying bifurcation but differ in how the bifurcation parameter evolves: one assumes a constant bifurcation parameter (steady-state null model), while the other assumes a linear change over time (test model). This approach has been studied, for example, in the context of fold bifurcations \cite{boettiger2012quantifying}. Similar procedures can also be applied to time-varying AR(p) models, in which autoregressive coefficients are either constant or allowed to vary over time \cite{dakos2012methods}.

Finally, other rank correlation coefficients could be employed to quantify trends. However, we expect that in general similar artifacts would be observed caused by the rolling windows methodology. We confirmed this by observing the same effect with the Pearson correlation coefficient (simulations not shown but available on request). We finally note that a better characterization of the asymptotic distribution of Mann-Kendall's tau could still serve as the basis for a dedicated statistical trend test for EWS detection. This would offer improved understanding and a faster alternative to computationally more intensive resampling-based methods.

\printbibliography



\section*{Data accessibility} 
The code for reproducing figures is accessible at \hyperlink{https://doi.org/10.5281/zenodo.19613302}{https://doi.org/10.5281/zenodo.19613302}.

\section*{Conflicts of Interest} 
The authors declare no conflicts of interest.

\section*{Acknowledgments}
The authors are very grateful to Michael Kopp and Jean-René Chazottes for helpful discussions and for providing useful comments on various versions of this manuscript.


\renewcommand\theequation{\Alph{section}\arabic{equation}} 
\counterwithin*{equation}{section} 
\renewcommand\thefigure{\Alph{section}\arabic{figure}} 
\counterwithin*{figure}{section} 
\renewcommand\thetable{\Alph{section}\arabic{table}} 
\counterwithin*{table}{section} 

\begin{appendices}

\section{The family of Mann-Kendall tests for trend testing}\label{annexe:kendall_tests}
\subsubsubsection{Asymptotic normality of Kendall's and Mann-Kendall's tau statistics}
Kendall proved that in the case of independent and identically distributed data, the asymptotic distribution of Kendall's $\tau$ is Gaussian \cite{kendall1975rank}. This is the following theorem: 
let's consider two sequences of random variables $X_i
\stackrel{iid}{\sim} X$ and $Y_i \stackrel{iid}{\sim} Y$, with $X$ and $Y$ two random variables and for $i \in \{1,...,n\}$. If $X$ and $Y$ are independent, then: 

\begin{equation*}
    \frac{\tau}{\sqrt{\mathbb{V}(\tau)}}\underset{n\to\infty}{\sim} \sqrt{\frac{9n}{4}} \tau \xrightarrow[n\to \infty]{d} \mathcal{N}(0,1)  
\end{equation*}

A similar result holds for Mann-Kendall's tau. Let's consider a sequence $X_i\stackrel{iid}{\sim} X$ for $i \in \{1,...,n\}$, then:
\begin{equation*}
     \frac{\tau_{\text{\tiny MK}}}{\sqrt{\mathbb{V}(\tau_{\text{\tiny MK}})}}\underset{n\to\infty}{\sim} \sqrt{\frac{9n}{4}} \tau_{\text{\tiny MK}} \xrightarrow[n\to \infty]{d} \mathcal{N}(0,1)  
\end{equation*}

\subsubsubsection{The original Mann-Kendall test}
Based on this asymptotic result, the Mann-Kendall test for trend is used to statistically assess the significance of the value of Mann-Kendall's tau and thus whether there is a monotonic upward or downward trend of the variable of interest over time \cite{mann1945nonparametric, kendall1975rank}. The Mann-Kendall null hypothesis for this test is that the $X_i$'s are independent and identically distributed. In this case, the variance of $S_{\text{\tiny MK}}$ and $\tau_{\text{\tiny MK}}$ are equal to:
\begin{align}\label{eq:V_independent}
    \mathbb V (S_{\text{\tiny MK}}) &= \frac{1}{18}n(n-1)(2n-5) \\
    \mathbb V (\tau_{\text{\tiny MK}}) &= \frac{\mathbb V (S_{\text{\tiny MK}})}{\binom{n}{2}^2} = \frac{2}{9}\frac{(2n-5)}{n(n-1)} = o_{n\rightarrow\infty}(1)
\end{align}
where $S_{\text{\tiny MK}} = \sum_{1\leq i<j \leq n} \textrm{sgn}(x_j-x_i)$ is a non-normalized quantity such that $\tau_{\text{\tiny MK}}   = \frac{S_{\text{\tiny MK}} }{\binom{n}{2}}$.

Thanks to this the asymptotic normality in the case of independent data, the standardized test statistic $Z$ is defined as:
\begin{equation} \label{eq:test_MK}
    Z= \left\{ 
    \begin{array}{ll}
        \frac{S_{\text{\tiny MK}}-1}{\sqrt{ \mathbb V (S_{\text{\tiny MK}})}} \text{ if } S_{\text{\tiny MK}}>0 \\
        0 \text{ if } S_{\text{\tiny MK}}=0 \\
        \frac{S_{\text{\tiny MK}}+1}{\sqrt{ \mathbb V (S_{\text{\tiny MK}})}} \text{ if } S_{\text{\tiny MK}}<0
    \end{array}
    \right.
\end{equation}
where $\pm 1$ is a continuity correction: $S_{\text{\tiny MK}}$ takes discrete values separated by $\pm 2$ (when two different observations cannot have the same value), thus it is a better approximation to consider that the tail of the distribution starts at $S_{\text{\tiny MK}}-1$ (resp. $S_{\text{\tiny MK}}+1$) if $S_{\text{\tiny MK}}$ is positive (resp. negative). Furthermore, the sign of $Z$ indicates an upward or downward trend. 
To assess the significance of the trend, the corresponding \textit{p-value} for the two-tailed test is: 
\begin{equation}\label{eq:p_value_MK}
    p = 1 - \sqrt{\frac{2}{\pi}}\int\limits_{0}^{\lvert Z\rvert} e^{-t^2/2}\mathrm{dt} = \sqrt{\frac{2}{\pi}}\int\limits_{\lvert Z\rvert}^{+\infty} e^{-t^2/2}\mathrm{dt}
\end{equation}

Thus, the construction of the Mann-Kendall test statistic and its tail probabilities fully relies on the Gaussian distribution of $\frac{S_{\text{\tiny MK}}}{\sqrt{\mathbb{V}(S_{\text{\tiny MK}})}}$ (or $\frac{\tau_{\text{\tiny MK}}}{\sqrt{\mathbb{V}(\tau_{\text{\tiny MK}})}}$ equivalently). This is always an approximation for finite time series but the rational for the use of this approximation is given by the fact that the asymptotic distribution of $\frac{S_{\text{\tiny MK}}}{\sqrt{\mathbb{V}(S_{\text{\tiny MK}})}}$ is Gaussian for i.i.d. data. And furthermore, Kendall proved that for time series of length $n$ greater than 10, the Gaussian approximation is already satisfactory \cite{kendall1975rank}. 

\subsubsubsection{Auto-correlated data} 
As mentioned above, the main assumption of the Mann-Kendall test is that the time series are sampled from independent and identically distributed random variables, which is often not the case in reality. 
For correlated random variables, the expected value of Mann-Kendall's tau is still zero (as $X_j-X_i$ has a symmetric distribution) but positive correlation increases its variance and thus the probability of rejecting the null hypothesis while negative correlation has the opposite effect and increases the probability of not rejecting \cite{yue2004mann}. 

\cite{bayley1946effective} has studied the variance of the mean $\Bar{X}$ of a sample of $n$ auto-correlated data from a random variable $X$ with variance $\sigma^2$. In this case, the variance of the mean is given by:
\begin{equation*}
    \mathbb V(\Bar{X}) = \frac{\sigma^2}{n} \frac{n}{n^*}  = \frac{\sigma^2}{n^*} 
\end{equation*}
where $n^*$ is the \textit{effective sample size} (\textit{ESS}) to take into account auto-correlation in the time series. Indeed, positive auto-correlation will result in an inflated variance of $\Bar{X}$ while negative auto-correlation will result in a reduced variance. $n^*$ is given by: 
\begin{equation}\label{eq:ESS}
    \frac{n}{n^*} = 1 + \frac{2}{n}\sum_{i=1}^{n-1}(n-i)\rho(i)
\end{equation}
where $\rho$ is the auto-correlation function of the data and $\rho(i)$ is the autocorrelation at lag-$i$ with $i\in\{1,...,n-1 \}$.

These results provide the rationale for applying the \textit{ESS} approach to correct statistics and eliminate the influence of auto-correlation. This was first proposed by \cite{lettenmaier1976detection} to correct the power of the parametric t-test and non-parametric Spearman’s rho and Mann-Whitney tests for auto-correlated time series. It is possible to follow the same procedure for Mann-Kendall's tau, defining and using a corrected variance to account for auto-correlation in data: 
\begin{equation*}
    \mathbb V^*(S_{\text{\tiny MK}}) = \mathbb V (S_{\text{\tiny MK}})\frac{n}{n_S^*} = \frac{n(n-1)(2n+5)}{18}\frac{n}{n_S^*}
\end{equation*}
where $\mathbb V (S_{\text{\tiny MK}})$ is the variance for independent data, cf. equation \eqref{eq:V_independent}. Although it is tempting to use the value of $\tfrac{n}{n^*}$ given in equation \eqref{eq:ESS}, \cite{hamed1998modified, yue2004mann} have shown that this gives very poor results. In particular, such a modified variance is a better approximation when there is no trend in the times series whereas when a trend is present, it influences the estimate of the auto-correlation function in equation \eqref{eq:ESS}. The trend, whether positive or negative, increases auto-correlation.

Two alternative methods have been suggested: one estimates auto-correlation after removing the trend and then uses equation \eqref{eq:ESS} \cite{yue2004mann}, while the other estimates auto-correlation on the ranks of the observations after removing the trend and uses a modified empirical version of equation \eqref{eq:ESS} \cite{hamed1998modified}. Then, for both proposed versions of the \textit{ESS}, the modified variance $\mathbb V^* (S_{\text{\tiny MK}})$ is used in the original Mann-Kendall trend test (presented in equations \eqref{eq:test_MK} and \eqref{eq:p_value_MK}) to take into account the auto-correlation in the data.

\subsubsubsection{Yue and Wang modified test for auto-correlated data}
Yue and Wang (\cite{yue2004mann}) have shown that if the \textit{ESS} defined in equation \eqref{eq:ESS} is computed with the sample auto-correlation which is estimated from the detrended time series, the \textit{ESS} can effectively reduce the influence of auto-correlation on the Mann-Kendall test.
It is therefore important to detrend the time series first, before estimating the auto-correlation. In the case where the existing trend can be approximated by a linear trend, the magnitude of the trend can be estimated using the non-parametric Theil-Sen estimator \cite{theil1950rank, sen1968estimates}. Non-linear trends must also be properly modeled and removed from time series first \cite{yue2004mann}. 

Furthermore, Yue and Wang only use the lag-1 coefficient $\rho_1$ to do their analysis in equation \eqref{eq:ESS}. Even though they do not justify this choice, it could be for the same reasons invoked in Hamed and Rao's method (presented in the next section): that is to disregard non-significant terms when calculating the \textit{ESS} so as not to have an effect on its value. 

\subsubsubsection{Hamed and Rao modified test for auto-correlated data}
Hamed and Rao (\cite{hamed1998modified}) also use the Theil-Sen estimator to detrend the time series, albeit without justification in this case. They then propose an empirical formula to modify the \textit{ESS} defined in equation \eqref{eq:ESS} thanks to the auto-correlation of the ranks of the observations: 
\begin{equation}\label{eq:ESS_Hamed}
    \frac{n}{n_S^*} = 1 + \frac{2}{n(n-1)(n-2)}\sum_{i=1}^{n-1}(n-i)(n-i-1)(n-i-2)\rho_{S_{\text{\tiny MK}}}(i)
\end{equation}
where $\rho_{S_{\text{\tiny MK}}}(i)$ is the auto-correlation function of the ranks of the data sequence. This is justified by the fact that Mann-Kendall's tau depends solely on the ranks between data and not on the values themselves. This method provides an approximation of the variance of Mann-Kendall's tau based on the auto-correlation between the ranks, which requires less computational effort.
However, as \cite{yue2002influence, yue2004mann} remark, although this method reduces the effect of auto-correlation on the Mann-Kendall test, rejection rates of the null hypothesis of no trend are still much higher than they should be. According to \cite{yue2004mann}, this could be due to the fact that the auto-correlation function on the ranks contains less information than the auto-correlation function itself.

Finally, as there is a large number of terms involved in the sum of equation \eqref{eq:ESS_Hamed}, \cite{hamed1998modified} only keeps $\rho_{S_{\text{\tiny MK}}}(i)$ values that are significant so that non-significant values do not have an effect on the variance approximation. This improves the variance approximation, but it also introduces a form of model selection potentially affecting rejection probabilities of the null hypothesis.

Furthermore, another type of test for auto-correlated data known as "exact" is also available in the literature \cite{hamed2009exact}. This test is based on the exact computation of the probabilities of all possible rankings of the Mann–Kendall trend test statistic for $n$ observations. However, it requires that the correlation structure of the process is perfectly known and not estimated (for example an AR(1) process with known parameter) and that the sample size $n$ is small (less than 9). Thus, it is not of much use in the general case and is not widely used.

\subsubsubsection{Normality of Mann-Kendall's tau in the case of auto-correlated data}
The two methods described above did not attempt to prove that Mann-Kendall's tau is approximately or asymptotically Gaussian when data are auto-correlated. However, based on numerical simulations, Hamed and Rao conclude that the normalized $S_{\text{\tiny MK}}$ is Gaussian regardless of whether the data are auto-correlated or not \cite{hamed1998modified}. Yue and Wang take this for granted too, and their modified test is based entirely on the Gaussian distribution of the statistic \cite{yue2004mann}. 

In the absence of formal proof in the case of auto-correlated data, other distributions were proposed over the years to better fit the distribution of $\tau_{\text{\tiny MK}}$. It was suggested that the generalized Gaussian distribution is a better approximation \cite{hu2020modified} while the Beta distribution was also proposed for moderate sizes of time series \cite{hamed2009exact}.

\section{Sensitivity analysis}\label{an:sensitivity_analysis}
In this appendix, we perform a sensitivity analysis with respect to several parameters defined in the main text. Our goal is to provide numerical evidence that the central results of the manuscript remain robust over a wide range of parameter values. Without loss of generality, we restrict the study to time series simulated from the fold normal form and Mann-Kendall statistic computed on the time series of lag-1 autocorrelation (the chosen EWS indicator), as Figure \ref{fig:similar_distributions} shows that the statistics are strongly influenced by $\alpha$, with minimal sensitivity to the bifurcation type or the EWS indicator chosen.

\subsection{Multiplicative noise}\label{an:multiplicative}
In the main text, we considered additive noise of the form: $$ \mathrm{d}x=f(x,r)\mathrm{d}t + \sigma \mathrm{d}W ,$$  
where $f$ is the normal form of local bifurcations presented in table \ref{tab:bif_liste} (parametrized by the bifurcation parameter $r$) and where $\sigma$ is independent of $x$. However, several EWS studies instead rely on multiplicative noise, typically of the form: 
$$\mathrm{d}x=f(x,r)\mathrm{d}t + \sigma x \mathrm{d}W .$$

\begin{figure}[!h]
    \centering
    \includegraphics[width=0.7\linewidth]{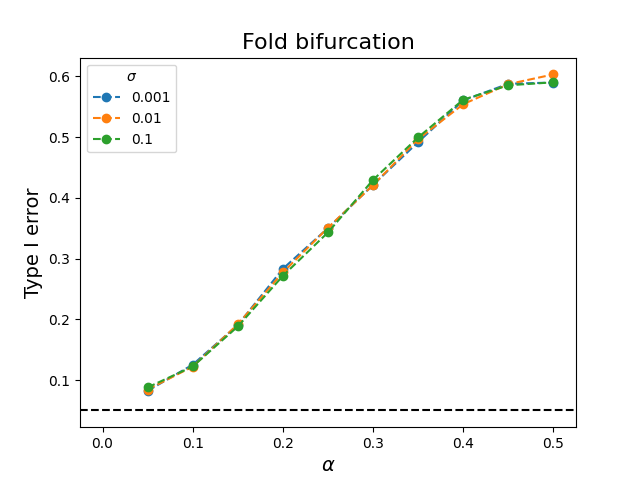}
    \caption{Empirical rejection rates of the null hypothesis of no trend carried out at a 5\% significance level for the Hamed and Rao modified Mann-Kendall test under multiplicative noise. The horizontal black line indicates the nominal 5\% significance threshold. Each original time series has length $N = 100$ and is simulated from the fold normal form (with fixed bifurcation parameter $r=-1$). The Mann-Kendall statistic is computed on the time series of lag-1 autocorrelation (the chosen EWS indicator), which is obtained from rolling windows of relative size $\alpha$ with stride $1$.}
    \label{fig:Appendix_mult_Noise}
\end{figure}

Figure \ref{fig:Appendix_mult_Noise} reproduces the analysis of Figure \ref{fig:window_test}  using this multiplicative noise structure. The rejection rates of the null hypothesis remain substantially above the expected 5\% significance level, confirming that the behaviour observed in the main text is not specific to the choice of additive noise.

\subsection{Noise intensity}
\begin{figure}[!h]
    \centering
    \includegraphics[width=0.7\linewidth]{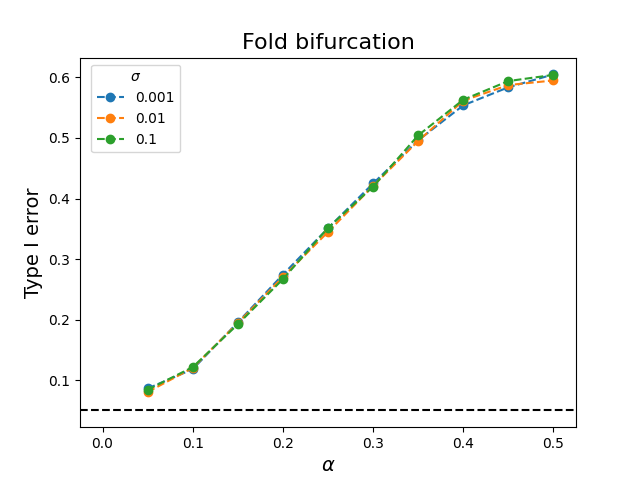}
    \caption{Empirical rejection rates of the null hypothesis of no trend at a 5\% significance level for the Hamed and Rao modified Mann-Kendall test for several values of the additive noise intensity $\sigma$. The horizontal black line indicates the nominal 5\% significance threshold. Each original time series has length $N = 100$ and is simulated from the fold normal form (with fixed bifurcation parameter $r=-1$). The Mann-Kendall statistic is computed on the time series of lag-1 autocorrelation (the chosen EWS indicator), which is obtained from rolling windows of relative size $\alpha$ with stride $1$.}
    \label{fig:Appendix_Noise}
\end{figure}

We also varied the noise intensity $\sigma$ of the additive noise used in the main text ($\sigma = 0.1$). Since our focus is on type I error rates - i.e., rejection rates under constant $r$ - the noise intensity solely controls the width of the distribution of the stochastic process that generates the original time series. As expected, this parameter has only a minor influence on type I error rates; see Figure \ref{fig:Appendix_Noise}.

\subsection{Sample size of the original time series}\label{an:length}
\begin{figure}[!h]
    \centering
    \includegraphics[width=0.7\linewidth]{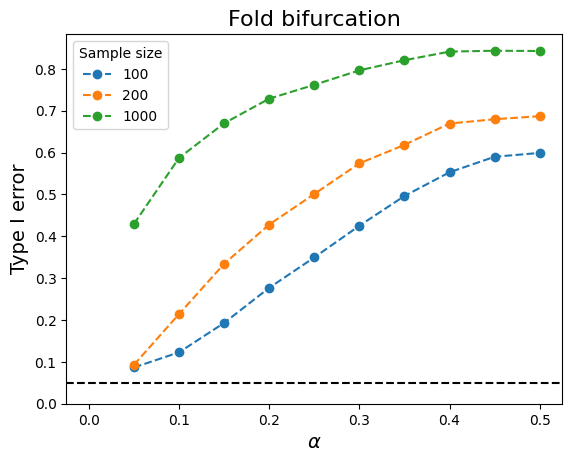}
    \caption{Empirical rejection rates of the null hypothesis of no trend at a 5\% significance level for the Hamed and Rao modified Mann-Kendall test for several sample sizes of the original time series. The horizontal black line indicates the nominal 5\% significance threshold. Each original time series is simulated from the fold normal form (with fixed bifurcation parameter $r=-1$). The Mann-Kendall statistic is computed on the time series of lag-1 autocorrelation (the chosen EWS indicator), which is obtained from rolling windows of relative size $\alpha$ with stride $1$.}
    \label{fig:Appendix_SampleSize}
\end{figure}

\begin{figure}[!h]
        \centering
    \begin{subfigure}[b]{1.\textwidth}
         \centering
         \includegraphics[width= 1.\textwidth]{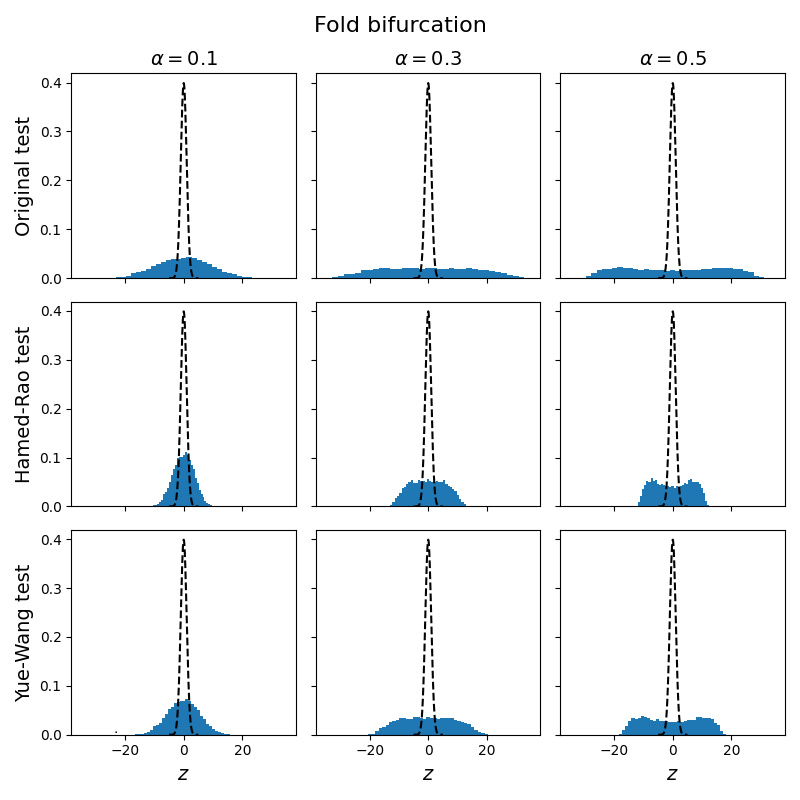}
     \end{subfigure}%
    \caption{Comparison of the empirical distributions of test statistics $Z$ of the original Mann-Kendall test (first row) the Hamed and Rao test (second row) and the Yue and Wang test (third row) with a standard normal distribution, the distribution hypothesized by the tests. The $Z$ statistic is the Mann-Kendall tau renormalized differently for each test and computed for the lag-1 autocorrelation coefficient time series estimated over rolling windows of relative size $\alpha$. Unlike Figure \ref{fig:distrib_X_rolling_windows} where the original time series is of length $N=100$, here $N=1000$. Time series are sampled from the fold bifurcation with fixed bifurcation parameter $r=-1$ (stationary null model).}
    \label{fig:an_N_sample}
\end{figure}

As discussed in the \textit{Discussion} section, increasing the sample size is expected to amplify the discrepancy between the empirical variance and the variance assumed by the tests, thereby inflating type~I error rates. Figure \ref{fig:Appendix_SampleSize} displays the resulting error rates as a function of the relative window size for several sample sizes ($N=100$ used in the main text). Moreover, Figure \ref{fig:an_N_sample} shows the empirical distribution of the modified Mann–Kendall statistic for $N=1000$, to be compared with Figure \ref{fig:distrib_X_rolling_windows} where $N=100$.

\subsection{Autocorrelation in the original time series}\label{an:autocorr}
\begin{figure}[!ht]
    \centering
    \includegraphics[width=0.7\linewidth]{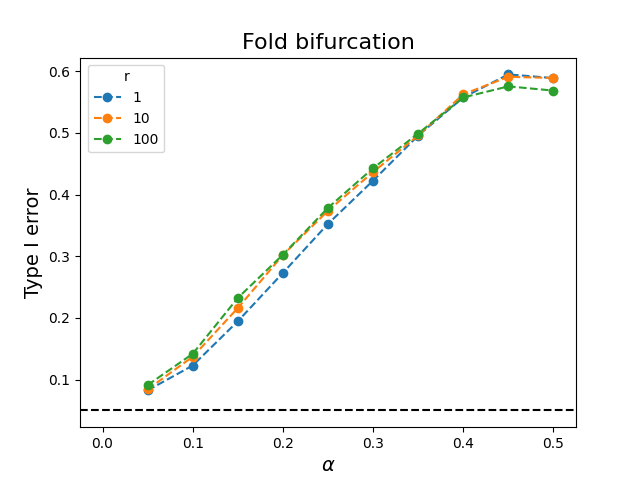}
    \caption{Empirical rejection rates of the null hypothesis of no trend at a 5\% significance level for the Hamed and Rao modified Mann-Kendall test, as a function of the relative rolling-window size $\alpha$, for several values of $\lvert r \rvert$. The horizontal black line indicates the nominal 5\% significance threshold. Each original time series has length $N = 100$ and is simulated from the fold normal form (with fixed bifurcation parameter $r=-1$). The lag-1 autocorrelation (the chosen EWS indicator) is computed using rolling windows of relative size $\alpha$ with stride $1$.}
    \label{fig:Appendix_Distance}
\end{figure}

Because the distance to the bifurcation, $\lvert r \rvert $, acts as a proxy for the process autocorrelation (see the autocorrelation function of the Ornstein-Uhlenbeck process appearing when linearizing), we also assessed how autocorrelation in the original time series may bias the modified Mann–Kendall tests. Figure \ref{fig:Appendix_Distance} shows the type I error rate as a function of $\alpha$ for different distances to the bifurcation. As $\lvert r \rvert $ increases, the original time series become progressively less autocorrelated; nevertheless, the type I error rates remain well above the nominal 5\% level, due to the overlap inherent in the rolling-window methodology.

\end{appendices}

\end{document}